\documentclass[aps,prd,reprint,twocolumn,10pt,eqsecnum,showpacs,nofootinbib,amsfonts,amssymb,amsmath,longbibliography,superscriptaddress,floatfix]{revtex4-2}
\usepackage[colorlinks=true,allcolors=blue, pdfusetitle]{hyperref}
\usepackage[all]{hypcap}
\usepackage{layouts}
\usepackage{bm}
\usepackage{mathrsfs}
\usepackage[dvipsnames]{xcolor}
\usepackage{mathtools}
\usepackage{graphicx}
\usepackage{orcidlink}
\usepackage{tensor}
\usepackage{subfigure}
\usepackage{printlen}
\graphicspath{{../plots/}{Paper_plots/}}

\usepackage[T1]{fontenc}
\usepackage{lmodern}
\usepackage[utf8]{inputenc}
\usepackage{microtype}

\DeclareMathAlphabet{\mathbfsf}{\encodingdefault}{\sfdefault}{bx}{sl}

\renewcommand{\Re}{\operatorname{Re}}
\renewcommand{\Im}{\operatorname{Im}}

\usepackage[normalem]{ulem}

\newcommand{\CornellPhysics}{\affiliation{Department of Physics, Cornell University, Ithaca, NY, 14853, USA}}
\newcommand{\Cornell}{\affiliation{Cornell Center for Astrophysics and Planetary Science, Cornell University, Ithaca, New York 14853, USA}}
\newcommand{\CornellLepp}{\affiliation{Laboratory for Elementary Particle Physics, Cornell University, Ithaca, New York 14853, USA}}
\newcommand{\Caltech}{\affiliation{Theoretical Astrophysics 350-17, California Institute of Technology, Pasadena, CA 91125, USA}}
\newcommand\UMiss{\affiliation{Department of Physics and Astronomy, University of Mississippi, University, MS 38677, USA}}
\newcommand\UMassD{\affiliation{Department of Mathematics, Center for Scientific Computing and Data Science Research, University of Massachusetts, Dartmouth, MA 02747, USA}}

\newcommand{\figparamspace}{%
\begin{figure}[t]
    \begin{center}
    \includegraphics[width=\columnwidth]{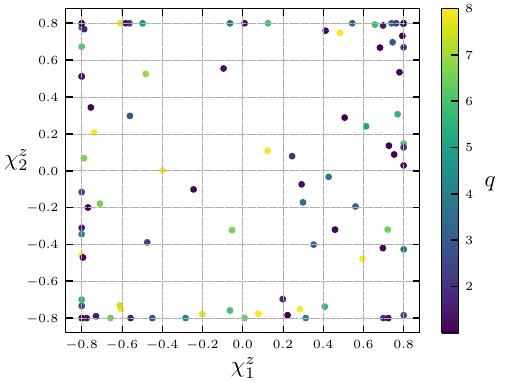}
    \end{center}
    \caption{Parameter space distribution of the input parameters $(q,\chi_{1}^{z},\chi_{2}^{z})$. The horizontal and vertical axes represent the dimensionless spin parameter magnitudes  of the two black holes and the color of each dot represents the mass ratio $q = \frac{m_{1}}{m_{2}}
    $ where $m_{1}$ is the larger mass of the two black holes.}
    \label{fig:paramspace}
  \end{figure}
}

\newcommand{\figsup}{%
\begin{figure*}[t]
    \begin{center} 
    \subfigure{ \includegraphics[width=\columnwidth]{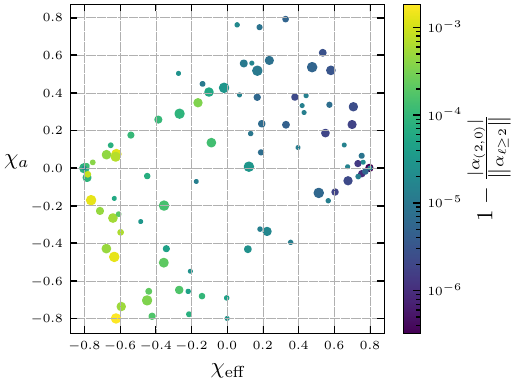} }
    \subfigure{ \includegraphics[width=\columnwidth]{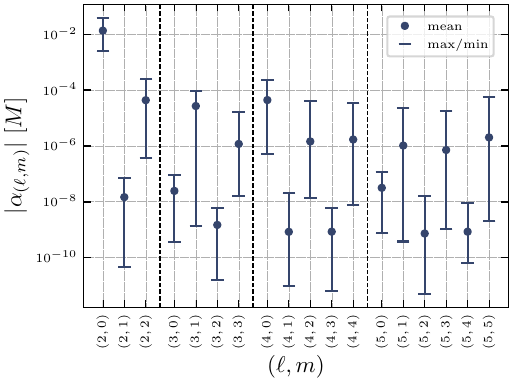} }
    \end{center}
    \caption{\emph{Left:} The relative difference of the $(2,0)$ mode amplitude with the $L^{2}$ norm $\Vert\alpha_{\ell\ge 2}\Vert$ of the proper supertranslation. The horizontal and vertical axes represent $\chi_{\text{eff}}$ [Eq.~\eqref{eq:chieff}] and $\chi_a$ [Eq.~\eqref{eq:chi_a}]. The dots' sizes represent the mass ratio $q$. Each dot represents one simulation from the catalog used. \emph{Right: } Modal composition (up to $\ell \le 5$) of the proper supertranslation dataset. The dots correspond to the dataset's mean for each mode amplitude, while the bars represent the minimum and maximum value over the dataset for each mode.}
    \label{fig:sup}
\end{figure*}
}

\newcommand{\figerrorsup}{%
\begin{figure*}[t]
    \begin{center}
    \subfigure{ \includegraphics[width=\columnwidth]  {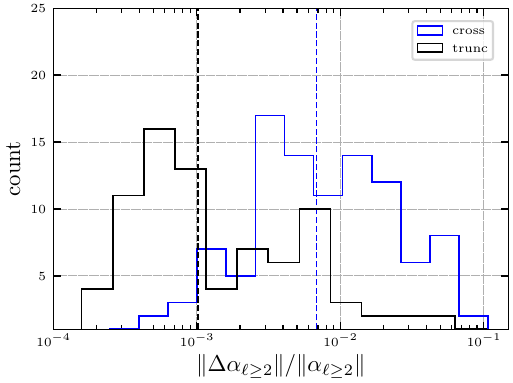} }
    \subfigure{ \includegraphics[width=\columnwidth]{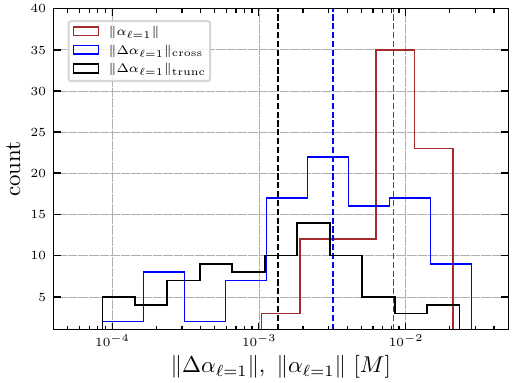} }
    \end{center}
    \caption{\emph{Left}: In blue we show the 5-fold cross-validation relative error of the surrogate model $\Vert\Delta\alpha_{\ell\ge 2}\Vert_{\text{cross}}/\Vert\alpha_{\ell\ge 2}\Vert$ and in black the relative truncation error $\Vert\Delta\alpha_{\ell\ge 2}\Vert_{\text{trunc}}/\Vert\alpha_{\ell\ge 2}\Vert$ of the proper supertranslation parameters of the dataset, obtained by feeding into the function $\Vert\Delta\alpha_{\ell\ge 2}\Vert$ [Eq.~\eqref{eq:propsuperr}] the data from the two highest simulation resolutions.
    \emph{Right:} In blue we show the 5-fold cross-validation error of the
    surrogate model $\Vert\Delta\alpha_{\ell=1}\Vert_{\text{cross}}$ and in
    black the truncation error $\Vert\Delta\alpha_{\ell=1}\Vert_{\text{trunc}}$
    of the space translation parameters of the dataset, obtained by feeding into
    the function $\Vert\Delta\alpha_{\ell=1}\Vert$ [Eq.~\eqref{eq:transerr}] the
    data from the two highest simulation resolutions. In red we show the space
    translation magnitudes $\Vert\alpha_{\ell=1}\Vert$ of the dataset. The
    histograms are not normalized and the vertical dashed lines represent the
    medians for each histogram.
}
    \label{fig:errorsup}
\end{figure*}
} 

\newcommand{\figerrorvel}{%
\begin{figure*}[t]
    \begin{center}
    \subfigure{ \includegraphics[width=\columnwidth]  {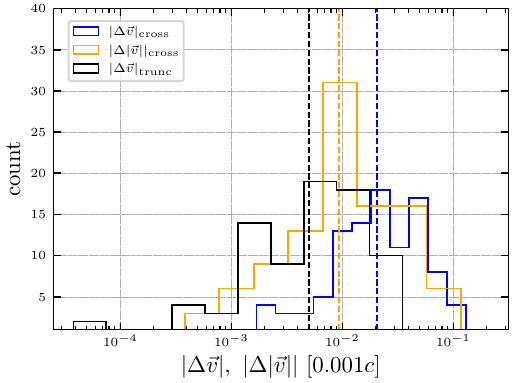} }
    \subfigure{ \includegraphics[width=\columnwidth]{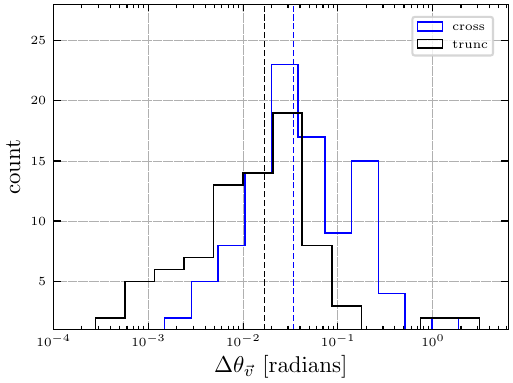} }
    \end{center}
    \caption{\emph{Left}: In blue and orange we show the 5-fold cross-validation errors of the surrogate model $|\Delta \vec{v}|_{\text{cross}}$,$|\Delta|\vec{v}||_{\text{cross}}$; in black we show the truncation error $|\Delta \vec{v}|_{\text{trunc}}$ of the boost parameters of the dataset, obtained by feeding into the function $|\Delta\vec{v}|$ [Eq.~\eqref{eq:normvelerr}] the data from the two highest simulation resolutions. The data is normalized by $[0.001c]$, which corresponds to the typical order of magnitude of the highest boosts in the dataset. 
    \emph{Right:} In blue we show the 5-fold cross-validation errors of the surrogate model $\Delta\theta_{\vec{v},\text{cross}}$. In black we show the truncation error $\Delta\theta_{\vec{v},\text{trunc}}$ of the boost parameters of the dataset, obtained by feeding into the function $\Delta\theta_{\vec{v}}$ [Eq.~\eqref{eq:angvelerr}] the data from the two highest simulation resolutions. The histograms are not normalized and the vertical dotted lines represent the medians for each histogram.
}
    \label{fig:magangvelerr}
\end{figure*}
} 

\newcommand{\figkickcomp}{%
\begin{figure*}[t]
    \begin{center} 
    \subfigure{ \includegraphics[width=\columnwidth]{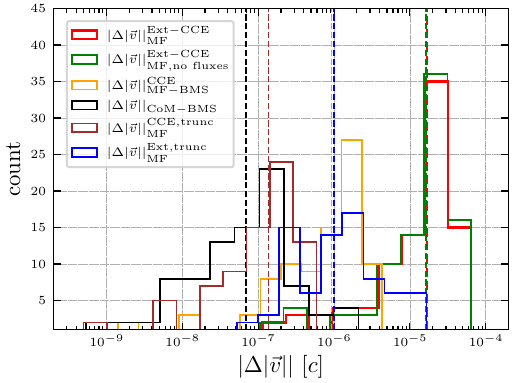} }
    \subfigure{ \includegraphics[width=\columnwidth]{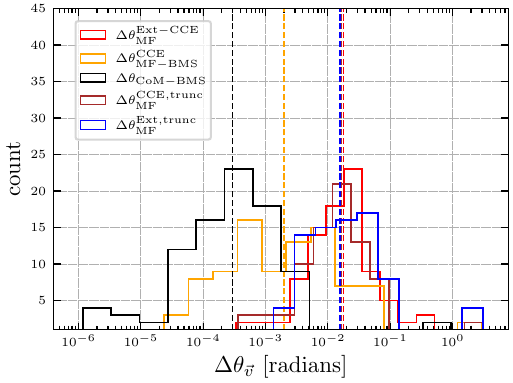} }
    \end{center}
    \caption{\emph{Left:} Absolute difference comparison between the dataset magnitudes of the momentum flux kicks $\vert\vec{v}_{\text{MF,CCE}}\vert$, $\vert\vec{v}_{\text{MF,Ext}}\vert$ [Eq.~\eqref{eq:momfluxkick}] on CCE and extrapolated waveforms, BMS boosts $\vert\vec{v}_{\text{BMS}}\vert$ and kicks obtained through Eq.~\eqref{eq:CoMKick} $\vert\vec{v}_{\text{CoM}}\vert$ (red, orange, black histograms). The green histogram shows the absolute difference of the dataset norms of the momentum flux kicks computed from CCE and extrapolated waveforms to which the energy and angular momentum fluxes have been subtracted from the strain. The brown and blue histograms show the truncation errors of the momentum flux kick norms for CCE and extrapolated waveforms, respectively. \emph{Right:} Relative angles between the dataset momentum flux kicks $\vert\vec{v}_{\text{MF,CCE}}\vert$, $\vert\vec{v}_{\text{MF,Ext}}\vert$ [Eq.~\eqref{eq:momfluxkick}] computed with CCE and extrapolated waveforms, BMS boosts $\vert\vec{v}_{\text{BMS}}\vert$ and kicks obtained through Eq.~\eqref{eq:CoMKick} $\vert\vec{v}_{\text{CoM}}\vert$ (red, orange, black histograms). The brown and blue histograms show the truncation errors of the relative angles between the CCE and extrapolated momentum flux kicks, respectively. The histograms are not normalized and the vertical dotted lines represent the medians for each histogram.}
    \label{fig:kickcomparison}
\end{figure*}
}
\newcommand{\figkickcomptwo}{%
\begin{figure}[t]
    \begin{center}
    \includegraphics[width=\columnwidth]{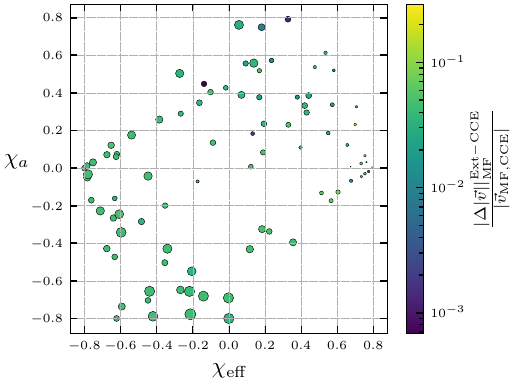}
    \end{center}
    \caption{Relative difference of the norms of momentum flux kicks evaluated on CCE waveforms and extrapolated waveforms of the dataset. Two symmetric systems with zero kick were not included in the plot. The horizontal and vertical axes represent $\chi_{\text{eff}}$ [Eq.~\eqref{eq:chieff}] and $\chi_a$ [Eq.~\eqref{eq:chi_a}]. The dots' size represents the CCE momentum flux kick magnitudes.}
    \label{fig:kickcomparison2}
  \end{figure}
}

\begin{document}

\author{Guido Da Re
  \orcidlink{0009-0007-2006-094X}}
\email{gdare@caltech.edu} \Caltech
\author{Keefe Mitman
  \orcidlink{0000-0003-0276-3856}}
\email{kem343@cornell.edu} \Cornell
\author{Leo C. Stein \orcidlink{0000-0001-7559-9597}} \UMiss
\author{Mark A. Scheel 
  \orcidlink{0000-0001-6656-9134}} \Caltech
\author{Saul A. Teukolsky~\orcidlink{0000-0001-9765-4526}} \Caltech\Cornell
\author{Dongze Sun
   \orcidlink{0000-0003-0167-4392}}
   \Caltech
\author{Michael Boyle \orcidlink{0000-0002-5075-5116}} \Cornell
\author{Nils Deppe \orcidlink{0000-0003-4557-4115}} \CornellLepp \CornellPhysics \Cornell
\author{Scott E. Field \orcidlink{0000-0002-6037-3277}} \UMassD
\author{Lawrence E.~Kidder \orcidlink{0000-0001-5392-7342}} \Cornell
\author{Jordan Moxon \orcidlink{0000-0001-9891-8677}} \Caltech
\author{Kyle C.~Nelli \orcidlink{0000-0003-2426-8768}} \Caltech
\author{William Throwe \orcidlink{0000-0001-5059-4378}} \Cornell
\author{Vijay Varma} \UMassD
\author{Nils L.~Vu \orcidlink{0000-0002-5767-3949}} \Caltech

\hypersetup{pdfauthor={Da Re et al.}}

\title{Modeling the BMS transformation induced by a binary black hole merger}

\begin{abstract}
Understanding the characteristics of the remnant black hole formed in a binary black hole merger is crucial for conducting gravitational wave astronomy. Typically, models of remnant black holes provide information about their mass, spin, and kick velocity. However, other information related to the supertranslation symmetries of the BMS group, such as the memory effect, is also important for characterizing the final state of the system. In this work, we build a model of the BMS transformation that maps a binary black hole's inspiral frame to the remnant black hole's canonical rest frame. Training data for this model are created using high-precision numerical relativity simulations of quasi-circular systems with mass ratios $q \le 8$ and spins parallel to the orbital angular momentum with magnitudes $\chi_{1}, \chi_{2} \le 0.8$. We use Gaussian Process Regression to model the BMS transformations over the three-dimensional parameter space $\left(q, \chi_{1}^{z}, \chi_{2}^{z}\right)$. The physics captured by this model is strictly non-perturbative and cannot be obtained from post-Newtonian approximations alone, as it requires knowledge of the strong nonlinear effects that are sourced during the merger. Apart from providing the first model of the supertranslation induced by a binary black hole merger, we also find that the kick velocities predicted using Cauchy-characteristic evolution waveforms are, on average, $\sim5\%$ larger than the ones obtained from extrapolated waveforms. Our work has broad implications for improving gravitational wave models and studying the large-scale impact of memory, such as on the cosmological background. The fits produced in this work are available through the Python package $\texttt{surfinBH}$ under the name $\texttt{NRSur3dq8BMSRemnant}$.
\end{abstract}
\maketitle
%
\section{Introduction}
\label{sec:introduction}
General relativity (GR) has always fascinated physicists and astronomers
for its extreme predictions, from the large-scale structure of the universe to incredibly
compact objects, such as black holes and neutron stars.
In 2015, a new way to test Einstein's theory emerged with the first detection of a gravitational
wave~\cite{LIGOScientific:2016aoc} by the laser interferometers LIGO (Laser Interferometer
Gravitational Wave Observatory) and Virgo.

Most of the gravitational radiation data collected so far comes from binary black hole mergers (BBHs). Thus, building precise theoretical models of such physical events is extremely important to perform stringent tests of general relativity. Since no analytic solution to the Einstein field equations is known for such systems, gravitational wave predictions can only be made by a joint endeavor from perturbation theory and numerical relativity (NR), which is the only tool available to explore the non-perturbative, strong gravity regime that characterizes the late inspiral, merger, and ringdown. 

One important aspect of modeling these phenomena is predicting the remnant black hole's properties. The knowledge of these parameters is crucial for constraining GR predictions, such as the no-hair theorem and the Kerr uniqueness theorem~\cite{Capano:2020dix,CalderonBustillo:2020rmh,LISAConsortiumWaveformWorkingGroup:2023arg,Isi:2021iql,Cotesta:2022pci,Bhagwat:2019dtm,Isi:2019aib,LIGOScientific:2021sio}, and performing merger population analyses~\cite{Doctor_2021,Mahapatra:2024qsy,Gupte:2024jfe,Yang:2019cbr,Gerosa:2021mno,mckernan2024mcfactsitestinglvk}. Furthermore, remnant properties are vital for black hole spectroscopy, as they determine the quasi-normal mode excitations during the ringdown. 

From the Kerr uniqueness theorem, the end state of a BBH is believed to be a Kerr black hole, fully characterized by its mass and spin. From the no-hair theorem, all the other degrees of freedom must be dissipated through gravitational radiation. Analyzing these events from large distances and at very early times (idealized as $t = -\infty$) and very late times (idealized as $t = +\infty$) enables them to be viewed as a scattering process~\cite{Varma:2018aht}, described by “in” and “out” states characterized by the initial spins and masses of the two black holes and the remnant black hole's properties. In the past, many models for the remnant's mass, spin, and recoil velocity as functions of the progenitor black holes' parameters have been proposed~\cite{Healy:2014yta,Zlochower:2015wga,Islam:2021mha,Thomas:2025rje,Varma:2018aht,Varma:2019csw}. 

Lately, improvements~\cite{Moxon:2020gha,Moxon:2021gbv} to the extraction of gravitational waves to future null infinity from numerical relativity simulations have made it possible to capture memory effects in BBHs, extending previous analytic and semi-analytical results~\cite{Zeldovich:1974gvh,Braginsky:1985vlg,Favata:2010zu,Christodoulou:1991cr,Christodoulou:1993uv} to a fully non-perturbative regime. These effects are associated with a net change in the structure of spacetime after the passage of a burst of gravitational radiation. Physically, the most prominent of these effects can be understood as the net displacement that two freely falling observers will experience after the passage of a gravitational wave transient. 

In the case of a BBH, these effects are mostly nonlinear and non-perturbative predictions of GR yet to be detected. These effects are linked through a generalized version of Noether's theorem~\cite{Wald:1999wa} to the asymptotic symmetry group of future null infinity, the BMS group~\cite{Bondi:1962px,Bondi1960,Sachs:1962wk,Sachs1962PR}. One may naively think that this group is simply some version of the Poincaré group, but it has been shown~\cite{Bondi:1962px,Bondi1960,Sachs:1962wk,Sachs1962PR} that it is an enlargement of this group that includes generalized angle-dependent spacetime translations called supertranslations. As waveforms are evaluated with respect to asymptotic inertial frames~\cite{MadlerWinicour2016} (or BMS frames) at null infinity, the BMS group governs all the transformations that map these frames into each other.

Nonlinear displacement memory can then be understood as a supertranslation from the canonical BMS frame of the inspiral to the BMS rest frame of the remnant black hole. Furthermore, the remnant kick velocity and net total angular momentum direction change can be described as a boost and rotation, completing a full BMS transformation between the two frames. Consequently, to fully characterize this black hole scattering process, it is crucial to model this BMS transformation, in addition to the other remnant parameters.

In this paper, we achieve this by modeling the BMS transformation from the inspiral to the remnant frame as a function of the progenitor spins and masses. The computations have been performed on 89 simulations associated with aligned-spin systems with mass ratios $q \le 8$ and dimensionless spin magnitudes $\chi_{1}$, $\chi_{2}\le 0.8$ using \texttt{SpECTRE}'s Cauchy-characteristic evolution (CCE) module~\cite{spectrecode,Moxon:2020gha,Moxon:2021gbv}. We build high-precision fits for the BMS transformation parameters over the 3-dimensional parameter space
$(q,\chi_{1}^{z},\chi_{2}^{z})$, using a non-parametric fitting algorithm called Gaussian Process Regression (GPR). The contribution of nonlinear displacement memory to the strain can be easily recovered from the supertranslation component of the modeled BMS transformation. Consequently, this model could play a relevant role for future GR tests that rely on gravitational wave memory, as these effects may become detectable with space-based interferometers such as LISA~\cite{Inchauspe:2024ibs,2022SCPMA..6519511Z}. Moreover, computing nonlinear memory from our model also opens up the possibility to refine estimations of the cosmological memory background from binary black hole mergers~\cite{2022SCPMA..6519511Z,Boybeyi_2024,Jokela_2022,zwick2024gravitationalwavememoryimprints}.

Our results could also improve the accuracy of existing waveform surrogate models~\cite{Yoo_2023} that include memory. Creating fits that accurately capture the features of both the inspiral and ringdown is challenging as they happen in different BMS frames. Our work suggests that one could build separate fits for the inspiral and ringdown and then stitch them together using our model of the BMS transformation between these two canonical frames. 

The paper is organized as follows. In Sec.~\ref{sec:BMSoverview} we briefly define the BMS group and the BMS charges used in the frame fixing procedure. In Sec.~\ref{sec:Numerical_evaluation} we show how we computed the BMS transformations for all the simulations in the dataset used. In Sec.~\ref{sec:surrogate} we describe how the GPR fitting procedure works and how it has been used to fit the parameters of our model. In Sec.~\ref{sec:errors} we discuss the accuracy of the fits and finally in Sec.~\ref{sec:Boostcomparison} we compare our results for the kick velocity of the remnant with other models.
\section{Overview of the BMS group}
\label{sec:BMSoverview}
The spacetime metric that describes an idealized binary black hole merger is asymptotically Minkowskian. After conformally compactifying the spacetime to include null infinity $\mathcal{I}^{+}$~\cite{Penrose1963, Penrose:1965am}, we can use the null asymptotic limit of Bondi coordinates $(u,r,\theta,\phi)$ as a chart on $\mathcal{I}^{+}$. This set of coordinates together with the associated Newman-Penrose tetrad on each tangent space on $\mathcal{I}^{+}$ physically represents an asymptotic inertial observer~\cite{MadlerWinicour2016}, or equivalently a \textit{BMS frame}.

The BMS group is the asymptotic symmetry group of asymptotically Minkowski spacetimes, and thus is the set of transformations that map BMS frames into each other while preserving the asymptotic form of the metric.\footnote{Formally, an asymptotic symmetry group is the quotient group of all the bulk gauge transformations that preserve the imposed asymptotic boundary conditions over all the trivial gauge transformation that reduce to the identity asymptotically. This quotient operation results in the definition of the asymptotic symmetry group of $\mathcal{I}^{+}$ for asymptotically flat spacetimes, called the BMS group.} One can naively think that this symmetry group is the Poincaré group, but as shown by Bondi, van der Burg, Metzer, and Sachs~\cite{Bondi:1962px,Bondi1960,Sachs:1962wk,Sachs1962PR} the actual group is an extension of this group that includes generalized angle-dependent spacetime translations called \textit{proper supertranslations}. For an in-depth review of the BMS group in relation to gravitational wave modeling, see Ref.~\cite{Mitman:2024uss}.

A general BMS transformation between two sets of inertial coordinates $(u,\theta,\phi)$ and $(u',\theta',\phi')$ on $\mathcal{I}^{+}$ can be written as:
\begin{align}
\begin{cases}
\label{eq:BMStrans_coord}
    u' = k(\theta,\phi)\left(u - \alpha(\theta,\phi)\right) \\
    (\theta',\phi') = f_{k,\boldsymbol{Q}}(\theta,\phi) ,
\end{cases}
\end{align}
where $f_{k,\boldsymbol{Q}}$ is a conformal transformation on the celestial sphere $S^2$ that depends on a unit quaternion $\boldsymbol{Q}$ and a boost conformal factor $k$
\begin{align}
\label{eq:conformalfactor}
k(\theta,\phi)\equiv\frac{\sqrt{1-|\vec{v}|^{2}}}{1-\vec{v}\cdot\hat{n}(\theta,\phi)}, 
\end{align}
where $\vec{v}$ is the boost velocity and $\hat{n}(\theta,\phi)$ a unit vector pointing in the $(\theta,\phi)$ direction on $S^{2}$. A supertranslation is described by the parameter $\alpha(\theta,\phi)$, a smooth function on the celestial sphere $S^2$, which can be decomposed into spherical harmonics as
\begin{align}
\label{eq:supdefinition}
\alpha(\theta,\phi) = \sum_{\ell=0}^{\infty}\sum_{\vert m\vert \le \ell} \alpha_{\ell m}Y_{\ell m}(\theta,\phi) .   
\end{align}
Thus, a BMS transformation is governed by 10 parameters characterizing Poincaré transformations and by infinitely many coefficients encoding proper supertranslations
\begin{align}
\begin{cases}
    \alpha_{00} & \text{time translations} \\
    \alpha_{1m} & \text{space translations} \\
    \vec{v} & \text{boosts} \\
    \boldsymbol{Q} & \text{rotations} \\
    \alpha_{\ell m} \quad \forall \ell \ge 2 &\text{proper supertranslations} .
\end{cases}
\end{align}

Since $\alpha$ is a real function, the spherical harmonics components $\alpha_{\ell m}$ have to satisfy
\begin{align}
\label{eq:alphareal}
  \alpha_{\ell m} = (-1)^{m}\bar\alpha_{\ell -m} , 
\end{align}
where an overbar denotes complex conjugation. Furthermore, $\alpha$ has the same units as $u$, which in this work has units of the total Christodoulou mass $M = m_1 + m_2$ of the binary~\cite{PhysRevLett.25.1596}.

Since BMS transformations in general do not commute with each other, in this work we decompose each element $g$ of the BMS group adopting the following composition order
\begin{equation}
    g = g_{\alpha} \circ g_{\boldsymbol{Q}} \circ g_{\vec{v}}
\end{equation}
where $g_{\alpha}$ is a supertranslation, $g_{\boldsymbol{Q}}$ a rotation, and $g_{\vec{v}}$ a boost.

Like a Poincaré transformation, a BMS transformation also acts on the tangent space by transforming the Newman-Penrose (NP) tetrad associated to the BMS frame at every point. This induces a transformation on the conformal Weyl scalars on $\mathcal{I}^{+}$ and thus on the gravitational strain as well~\cite{Bondi:1962px,Sachs:1962wk,NewmanUnti1962,Boyle2016}. It is clear then that a correct choice of the BMS frame is necessary if one wants to extract correct physical properties from a waveform, or compare waveforms computed in different frameworks.

It can be shown~\cite{deBoer:2003vf,Banks:2003vp,Barnich:2009se,Barnich:2010ojg} that through a generalized version of Noether's theorem, conservation laws can be written for each generator of the BMS group. These conservation laws relate the BMS symmetry charges to their fluxes and to memory effects. If we look at these balance laws in non-radiative sections of $\mathcal{I}^{+}$, such as at very early times ($u= -\infty$) and at very late times (when the spacetime has relaxed down to the remnant black hole), then we can quantify memory effects and the remnant kick as due to the difference of the BMS charges measured between the two canonical BMS frames at $u = \pm \infty$.

The waveform data produced by numerical relativity simulations is in some arbitrary numerical BMS frame, resulting from how the initial data is chosen and whatever the evolution gauge is. We will call this frame the NR BMS frame.

To map to the remnant frame we simply need to have the BMS charges of the NR system match the values of the ones associated with the stationary Kerr metric.
The inspiral frame at $u = -\infty$ is the same as the post-Newtonian frame, thus we can use the BMS charges associated with a PN system as a proxy to map into such a frame.

The boost velocity and spacetime translation are fixed by minimizing a linear fit to the center-of-mass charge over a time window $\Delta u$ around the mapping time $u^{*}$. This charge is defined as~\cite{Kozameh:2013bha,Compere:2019gft,Mitman:2022kwt}
\begin{align}
\label{eq:comcharge}
G^{a}(u)&=(K^{a}+uP^{a})/P^{t}\nonumber\\
&=\frac{1}{4\pi}\int_{S^{2}}\mathrm{Re}\left[\left(\bar{\eth}n^{a}\right)\left(\hat{N}+u\eth m\right)\right]\,d\Omega/P^{t},
\end{align}
where $P^{a}$ is the Bondi four-momentum, $K^{a}$ is the boost charge~\cite{GomezLopez:2017kcw,Dray:1984rfa,Dray:1984gz,Streubel1978}, $n^{a}$ is a null unit normal in the $(\theta,\phi)$ direction, $m$ is the Bondi mass aspect and $\eth$ is the spin-weight raising operator~\cite{Geroch:1973am}.

The rotation is fixed by looking at two different charges for the inspiral and remnant frame. In the inspiral case we look at the angular velocity directly defined from the strain (or news) modes as~\cite{Boyle:2013nka}
\begin{align}
\label{eq:angularvelocity}
\vec{\omega}(u)=-\langle\vec{L}\vec{L}\rangle^{-1}\cdot\langle\vec{L}\partial_{t}\rangle,
\end{align}
where
\begin{subequations}
\begin{align}
\langle\vec{L}\partial_{t}\rangle^{a}&\equiv\sum\limits_{\ell,m,m'}\mathrm{Im}\left[\bar{f}_{(\ell,m')}\langle\ell,m'|L^{a}|\ell,m\rangle\dot{f}_{(\ell,m)}\right],\\
\langle\vec{L}\vec{L}\rangle^{ab}&\equiv\sum\limits_{\ell,m,m'}\bar{f}_{(\ell,m')}\langle\ell,m'|L^{(a}L^{b)}|\ell,m\rangle f_{(\ell,m)}.
\end{align}
\end{subequations}
The operator $\vec{L}$ is
the infinitesimal generator of rotations
and $f(u,\theta,\phi)$ is some function corresponding to the
asymptotic radiation, such as the strain or the news. In particular, we can ask
this vector to be the same as the one evaluated from a PN waveform at the
target mapping time $u^{*}$. The PN waveforms used for this project are created using the python package GWFrames~\cite{GWFrames}. The PN orbital phase includes non-spinning terms
up to 4 PN order~\cite{Blanchet:2004ek,Blanchet:2013haa,Jaranowski:2013lca,Bini:2013zaa,Bini:2013rfa} and spinning terms up to 2.5
PN order~\cite{Kidder:1995zr,Will:1996zj,Bohe:2012mr}. The PN amplitude includes
non-memory terms to 3.5 PN order~\cite{Blanchet:2008je,Faye:2012we,Faye:2014fra}, non-spinning
memory terms to 3 PN order, and spinning memory terms
up to 2 PN order~\cite{Favata:2008yd,Mitman:2022kwt}. We use the TaylorT4~\cite{Boyle:2007ft}
approximant to compute the PN phase.

For the remnant black hole case, as we do not have
orbital angular momentum contributions we instead use the asymptotic
dimensionless spin vector~\cite{Iozzo:2021vnq}
\begin{align}
    \vec{\chi}(u) = \frac{\gamma}{M_{B}^2}\left(\vec{J} + \vec{v} \times \vec{K}\right) - \frac{\gamma -1}{M_{B}^2}\left(\hat{v}\cdot \vec{J}\right)\hat{v},    
\end{align}
where
\begin{align}
    \gamma(u) = 1/\sqrt{1 - \vert \vec{v}^2\vert}
\end{align}
is the Lorentz factor,
\begin{align}
    M_{B}(u) \equiv \sqrt{-\eta_{ab}P^{a}P^{b}}   
\end{align}
is the Bondi mass,
\begin{equation}
    \vec{v}(u) \equiv \vec{P} /P^t 
\end{equation}
is the velocity vector, and $\vec{J}$ is the angular momentum Poincaré charge~\cite{Mitman:2022kwt}.
This completely fixes the rotation freedom in both frames up to a constant $U(1)$ phase rotation. For mapping to the PN BMS frame, this rotation freedom is fixed by a time and phase alignment that minimizes the mismatch between the NR and PN waveforms. For mapping to the remnant BMS frame the rotation freedom is fixed by preserving the $x$-axis's orientation relative to the original frame.

Finally, the supertranslation freedom is fixed by using another charge, called the Moreschi supermomentum~\cite{OMMoreschi_1986,Moreschi:1988pc,Moreschi:1998mw,Dain:2000lij}, which generalizes the Bondi four-momentum
\begin{align}
\label{eq:Moreschisupermomentum}
\mathcal{P}_{\mathrm{M}}(u,\theta,\phi) &= \Psi_{2} + \sigma\dot{\bar{\sigma}} + \eth^{2}\bar{\sigma} \nonumber\\
& = \int_{-\infty}^{u}|\dot{\sigma}|^{2}\,du-M_{\mathrm{ADM}},
\end{align}
where $\sigma$ is the shear (equivalent to the strain $h$ up to a constant set by the normalization convention on the NP tetrad), $M_{\mathrm{ADM}}$ the ADM mass and $\Psi_{2}$ one of the asymptotic Weyl scalars. This charge can be physically thought of as the energy radiated up to the time $u$ by the binary in the direction $(\theta,\phi)$, which through the supertranslation balance law, sources the instantaneous nonlinear displacement memory.

The supertranslation $\alpha$ that maps $\mathcal{P}_{\mathrm{M}}$ to some target value is found through a procedure first defined in Ref.~\cite{Dain:2000lij} and numerically implemented in Ref.~\cite{Mitman:2021xkq}, which starts by looking at how the Moreschi supermomentum transforms under a BMS transformation
\begin{align}
    \mathcal{P}_{\mathrm{M}}'=\frac{1}{k^{3}}(\mathcal{P}_{\mathrm{M}}-\eth^{2}\bar{\eth}^{2}\alpha),
\end{align}
where $\mathcal{P}_{\mathrm{M}}'$ is the supermomentum in the target frame, and $k$ is the conformal factor of the boost. The supertranslation $\alpha$ defines a section $u = \alpha$ on $\mathcal{I}^{+}$, on which the value of the transformed supermomentum is $\mathcal{P}_{M}'$. As a consequence we can write
\begin{align}
\label{eq:nicesectioneq}
    \eth^{2}\bar{\eth}^{2}\alpha = \mathcal{P}_{\mathrm{M}}(\alpha,\theta,\phi) - \mathcal{P}_{\mathrm{M}}'(\alpha,\theta,\phi) k^{3}(\alpha,\theta,\phi) .
\end{align}
The existence and uniqueness of the solutions of this equation was proved in Ref.~\cite{Dain:2000lij}. The solution is extracted numerically through an iterative process, starting with the initial guess $\alpha = u_{*}$, as described in Ref.~\cite{Mitman:2021xkq}.

In order to map to the PN BMS frame, one can simply ask $\mathcal{P}_{M}'(\alpha,\theta,\phi)$ to be equal to the PN Moreschi supermomentum, solving Eq.~\eqref{eq:nicesectioneq} for some initial guess $u^{*}_{\text{PN}}$ in the inspiral. The PN Moreschi supermomentum has been semi-analytically calculated up to 3PN order for non-spinning black holes and 2PN order for spinning black holes with no eccentricity~\cite{Mitman:2022kwt}.
After the ringdown the system approaches a stationary black hole, thus we choose the remnant BMS frame to be shear-free. This results in the condition $\mathcal{P}_{\mathrm{M}}'(u = \alpha, \theta,\phi) = \Psi_{2}' (\alpha, \theta,\phi) = - M_{B} (\alpha)$, where $M_{B}$ is the Bondi mass of the remnant black hole, and Eq.~\eqref{eq:nicesectioneq} is solved for a post-ringdown initial guess $u^{*}_{\mathrm{Rem}}$.

After this procedure, all the infinitely many parameters that characterize the desired BMS transformation are fixed.
\section{Numerical evaluation of the BMS transformation}
\label{sec:Numerical_evaluation}
With the frame fixing procedure clarified, we now explain how the BMS transformation from the inspiral frame to the remnant frame is computed.

Call $g_{\text{Rem}}$ the BMS transformation that maps the NR BMS frame to the remnant black hole BMS frame and $g_{\text{PN}}$ the BMS transformation that maps the NR BMS frame to the PN BMS frame. We can calculate the BMS transformation from the inspiral to the remnant frame as
\begin{align}
    g_{\text{Insp} \to \text{Rem}} = g_{\text{Rem}} \circ g_{\text{PN}}^{-1} ,
    \label{eq:basiccomprule}
\end{align}
where $\circ$ denotes group composition. Furthermore, to address the additional
phase fixing gauge freedom in the orbital plane, we rotate the inspiral frame to
have the two black holes aligned to the $x$ axis $100 M$ before the merger time
$u_{\text{peak}}$. To fix the remaining $\pi$ phase freedom, we choose to have
the heaviest black hole on the positive-$x$ axis. This ensures the same phase
fixing choice as Ref.~\cite{Varma:2018aht}. The merger time is chosen to be the
one that maximizes the $L^{2}$ norm of the news $\dot{h}$\footnote{Note that in Ref.~\cite{Varma:2018aht} the merger time is calculated from the strain and not from the news, this choice is not significant for the results of this work.}
\begin{align}
   u_{\text{peak}} = \underset{u}{\mathrm{argmax}}\ \Vert \dot{h}\Vert.
\end{align}
These conditions are achieved by looking at the coordinate positions of the black holes' apparent horizons with respect to the center-of-mass in the simulation domain (Cauchy domain). The coordinates of the apparent horizons' centers are related to the NR BMS frame. Thus in order to use this information in the PN BMS frame we have to rotate the position vectors of the two black holes by the same rotation that the system undergoes when transforming to the PN BMS frame. 

Defining $u_{0} = u_{\text{peak}} - 100 M$, we can define $\boldsymbol{r}_{A,B}(u_{0})$ as the coordinates of the apparent horizon centers in the Cauchy domain's inertial frame at $u_{0}$. Given the unit quaternion $\boldsymbol{Q}_{\text{PN}}$, associated to the frame rotation from the NR BMS frame to the PN BMS frame, we can apply this rotation to $\boldsymbol{r}_{A,B}(u_{0})$, which in quaternion notation gives
\begin{equation}
    \boldsymbol{r}'_{A,B} = \boldsymbol{Q}_{PN}^{-1}\boldsymbol{r}_{A,B}\boldsymbol{Q}_{PN} .
\end{equation}
The Newtonian center-of-mass associated with the centers of the apparent horizons is
\begin{align}
\label{eq:22phasefix}
    \boldsymbol{r}'_{\mathrm{CoM}}(u_{0}) = \frac{m_{A}\boldsymbol{r}'_{A}(u_{0}) + m_{B}\boldsymbol{r}'_{B}(u_{0})}{m_{A} + m_{B}} ,
\end{align}
where $m_{A} \ge m_{B}$ are the Christodoulou masses of the two black holes. The unit normals of the position vectors of the two black holes relative to $\boldsymbol{r}'_{\mathrm{CoM}}(u_{0})$ are simply:
\begin{align}
    \Delta \hat{\boldsymbol{r}}'_{A}(u_{0}) = \frac{\boldsymbol{r}'_{A}(u_{0}) - \boldsymbol{r}'_{\mathrm{CoM}}(u_{0})}{\vert\boldsymbol{r}'_{A}(u_{0}) - \boldsymbol{r}'_{\mathrm{CoM}}(u_{0})\vert} = -  \Delta \hat{\boldsymbol{r}}'_{B}(u_{0}) .
\end{align}
Now we can define a unit quaternion $\boldsymbol{Q}$ such that in quaternion notation:
\begin{align}
    \hat{\boldsymbol{x}} = \boldsymbol{Q}\Delta \hat{\boldsymbol{r}}'_{A}(u_{0})\boldsymbol{Q}^{-1} = - \boldsymbol{Q}\Delta \hat{\boldsymbol{r}}'_{B}(u_{0})\boldsymbol{Q}^{-1},
\end{align}
where $\hat{\boldsymbol{x}}$ is the unit vector of the $x$-axis.

We implement this rotation $\boldsymbol{Q}$ in the composition rule by modifying Eq.~\eqref{eq:basiccomprule} to\footnote{Note that the BMS transformation~\eqref{eq:fullbmscomp} can be achieved equivalently by performing first $g_{\text{PN}}$, followed by $\boldsymbol{Q}$ on the waveform data and from that BMS frame fix directly to the remnant frame, obtaining then $g_{\text{Insp} \to \text{Rem}} = g_{\text{PN},\boldsymbol{Q} \to \text{Rem}}$.}
\begin{align}
\label{eq:fullbmscomp}
    g_{\text{Insp} \to \text{Rem}} = g_{\text{Rem}} \circ g_{\text{PN}}^{-1} \circ g_{\boldsymbol{Q}}^{-1} ,
\end{align}
where $g_{\boldsymbol{Q}}$ is the element of the BMS group that corresponds to the unit quaternion $\boldsymbol{Q}$.

For each simulation, the waveform data and Weyl scalars are extracted with the \texttt{SpECTRE} code's CCE module~\cite{spectrecode,Moxon:2020gha,Moxon:2021gbv}. Note that CCE, unlike methods based on extrapolation~\cite{Boyle_2009,Boyle:2019kee}, can correctly reproduce memory by solving Einstein equations on null slices connecting the Cauchy domain to future null infinity.
The numerical handling of waveform data, BMS transformations, and frame fixing has been carried out using the python package \texttt{scri}~\cite{scri, Boyle:2013nka, Boyle:2014ioa,Boyle2016}. 

This procedure has been applied to 89 aligned-spin simulations with $q \le 8$
and $\vert\chi_{1,2}^{z}\vert \le 0.8$, where $\chi_{1,2}^{z}$ denotes the $z$
component of the dimensionless spins of the two black holes. The simulation
dataset includes a subset of simulations from the SXS catalog~\cite{Boyle:2019kee},
numbered SXS:BBH:1419 to SXS:BBH:1510, excluding SXS:BBH:1468 and SXS:BBH:1488\footnote{We exclude these two simulations because the worldtubes required for running CCE were not available at the start of this project.}.
The input parameter space $(q, \chi_{1}^{z}, \chi_{2}^{z})$ can be seen in
Fig.~\ref{fig:paramspace}.
\figparamspace

For the PN BMS frame fixing we choose to map at a time $u^{*}_{\text{PN}}$ that corresponds to 12 orbits before the merger with a window $\Delta u_{\text{PN}}$ corresponding to 2 orbits, while to map to the remnant frame we set the mapping time $u^{*}_{\text{Rem}}$ to be $20 M$ before the end of the waveform with a window length $\Delta u_{\text{Rem}} = 20 M$. This ensures that the frame fixing procedure is performed when the loudest quasi-normal modes of the ringdown are well decayed and the spacetime is close to the stationary state of the remnant black hole. 

Out of the complete set of supertranslation parameter modes, we model only the space translations and the proper supertranslations
\begin{align}
    \alpha_{\ell = 1}(\theta,\phi) &= \sum_{\vert m\vert \le 1} \alpha_{1 m}Y_{1 m}(\theta,\phi) \,,\\
    \alpha_{\ell \ge 2}(\theta,\phi) &= \sum_{\ell \ge 2}^{\ell_{\text{max}}}\sum_{\vert m\vert \le \ell} \alpha_{\ell m}Y_{\ell m}(\theta,\phi) ,
\end{align}
as the influence of time translation solely results in a uniform shift in time for each mode of any asymptotic quantity, thus carrying no physical meaning. In this work we model the modes up to $\ell_{\text{max}} = 8$.

The proper supertranslation amplitudes obtained are larger for lower mass ratios and for systems having larger spins pointing in the direction of the orbital angular momentum, in agreement with other results in the literature characterizing the displacement memory's dependence on the input parameters~\cite{Inchauspe:2024ibs,Favata:2008yd,2022SCPMA..6519511Z,PhysRevD.98.064031,PhysRevD.103.043005,Cao_2016,Pollney_2011}.

The left panel of Fig.~\ref{fig:sup} shows the relative contribution of the $(2,0)$ mode amplitudes of $\alpha$ in the dataset over the total $L^{2}$ norm $\Vert\alpha_{\ell\ge 2}\Vert$ in terms of $q$, $\chi_{\mathrm{eff}}$, and $\chi_{a}$, where
\begin{align}
    \Vert\alpha_{\ell\ge 2}\Vert = \sqrt{\sum_{\ell \ge 2}^{\ell_{\text{max}}}\sum_{\vert m\vert \le \ell} \vert\alpha_{\ell m}\vert^{2}} \,,
\end{align}
and
\begin{align}
\label{eq:chieff}
    &\chi_{\mathrm{eff}} \equiv \frac{q\chi_{1}^{z} + \chi_{2}^{z}}{1+q},\\
    \label{eq:chi_a}
    &\chi_{a}\equiv \frac{\chi_{1}^{z}-\chi_{2}^{z}}{2} .
\end{align}
We can clearly see that the $(2,0)$ mode contributes to most of the proper supertranslation content, as expected since the Moreschi supermomentum $\mathcal{P}_{M}$ is largely dominated by the $(2,0)$ mode. Consequently, that mode contributes the most to build the nonlinear displacement memory in the full strain. At lower $\chi_{\mathrm{eff}}$ and $\chi_{a}$ values we see that other modes start to contribute more, but the $(2,0)$ mode remains largely dominant. 

In the right panel of Fig.~\ref{fig:sup} we show the spectrum of each mode amplitude minimum, maximum and mean values over the dataset. We can see that the modes with $\ell + m$ even are largely dominant with respect to the odd ones. This is because modes with $\ell + m$ even have electric parity.

The boost parameter $\vec{v}$'s dataset features are discussed in more detail in Sec.~\ref{sec:Boostcomparison}. 

Finally, we note that the rotation parameters cannot be modeled by GPR. As the system is non-precessing, the final black hole has its spin
always aligned with the $z$ direction, but its orientation could differ from the
inspiral total angular momentum orientation up to a $\pi$ flip about the orbital
plane. As the frame rotation describes this direction flip, modulo a $U(1)$ rotation in the orbital plane, this transition cannot be modeled by any smooth function and thus cannot be fit with GPR. Therefore, since the rotation parameter is not modeled, the remnant BMS frame and the inspiral BMS frame share the same $U(1)$ gauge choice.
We note though that for precessing systems
fitting the
rotation parameters is meaningful, as continuous spin orientations of the remnant
black holes are possible.
\figsup
\section{Building a surrogate model on the transformation parameters}
\label{sec:surrogate}
Numerical relativity simulations are expensive calculations that require a large amount of time to run. Consequently, producing waveform or remnant models that cover the progenitor parameter space poses a challenge. An approach to overcome this limitation is called surrogate modeling. The key concept is using a machine learning-based algorithm trained directly on numerical relativity simulations to predict data for any input parameters inside the training range. In Ref.~\cite{Varma:2018aht} Gaussian Process Regression (GPR) was used to build a set of surrogate models for the properties of the remnant black hole in a BBH, in particular the final mass, spin, and kick velocity. 

In this work we build a surrogate model following their procedure, namely using GPR to fit the supertranslation parameters and the boost velocities in terms of the input parameters $(q,\chi_{1}^{z},\chi_{2}^{z})$. GPR is a Bayesian, non-parametric fitting method that relies on Gaussian processes to model the shape of a function, based on the input dataset~\cite{b60dec2e1b6c416387f33e9de784f573,Wang_2023,Varma:2018aht}. The only assumption made \emph{a priori} is that the function that we want to model can be described as a Gaussian process with some smoothness features captured by the choice of a suitable kernel. 

In particular, given a function of the form $f(\boldsymbol{x})$, with $\boldsymbol{x}$ an input vector of dimension $D$, the prior distribution of the possible functions that we can model is given by a multivariate normal distribution. More precisely, for any finite subset of input points $\vec{\boldsymbol{x}}$\footnote{From now on a bold character means a vector in the parameter space, while the top arrow sign denotes the vector of all the inputs or outputs of a given dataset. Equivalently, the arrow notation can be replaced by an index notation (latin indices) where needed.} we have~\cite{b60dec2e1b6c416387f33e9de784f573,Wang_2023,Varma:2018aht}
\begin{align}
     \mathcal{P}(\vec{f}|\vec{\boldsymbol{x}}) = \mathcal{N}(\vec{m},K_{\boldsymbol{x}\boldsymbol{x}}) ,    
\end{align}
where $[\vec{m}]_{i} = m_{i} = m(\boldsymbol{x}_{i})$ is the mean function (usually set to 0), $[\vec{f}]_{i} = f_{i}$ are the outputs and $[K_{\boldsymbol{x}\boldsymbol{x}}]_{ij} = K(\boldsymbol{x}_{i},\boldsymbol{x}_{j})$ is the covariance matrix of the distribution, also called the kernel. Following this assumption, the training dataset $\{\boldsymbol{x}_{i},f(\boldsymbol{x}_{i})\}$ with $i=1,\dots,N$ and the set of predictions that we want to make $\{\boldsymbol{x}^{*}_{j},f^{*}(\boldsymbol{x}^{*}_{j})\}$ with $j=1,\dots,M$ must also follow the same multivariate normal distribution, or joint gaussian distribution~\cite{b60dec2e1b6c416387f33e9de784f573,Wang_2023,Varma:2018aht}
\begin{align}
    \mathcal{P}(\vec{f},\vec{f}^{*}| \vec{\boldsymbol{x}},\vec{\boldsymbol{x}}^{*}) = \mathcal{N}\left(0,\begin{bmatrix}
        K_{\boldsymbol{x}\boldsymbol{x}} & K_{\boldsymbol{x}\boldsymbol{x}^{*}} \\
        K_{\boldsymbol{x}^{*}\boldsymbol{x}} & K_{\boldsymbol{x}^{*}\boldsymbol{x}^{*}}
    \end{bmatrix}\right) .  
\end{align}
Then from Bayes theorem the conditional probability distribution of the predictions given the inputs and the training dataset is
\begin{align}
    \mathcal{P}(\vec{f}^{*}| \vec{f},\vec{\boldsymbol{x}},\vec{\boldsymbol{x}}^{*}) = \frac{\mathcal{P}(\vec{f},\vec{f}^{*}| \vec{\boldsymbol{x}},\vec{\boldsymbol{x}}^{*})}{\mathcal{P}(\vec{f}|\vec{\boldsymbol{x}})}.
\end{align}
From the properties of multivariate gaussian distributions and with some algebra~\cite{b60dec2e1b6c416387f33e9de784f573,Wang_2023} we get
\begin{align}
    \mathcal{P}(\vec{f}^{*}| \vec{f},\vec{\boldsymbol{x}},\vec{\boldsymbol{x}}^{*}) = \mathcal{N}(K_{\boldsymbol{x}^{*}\boldsymbol{x}}K_{\boldsymbol{x}\boldsymbol{x}}^{-1}\vec{f}, K_{\boldsymbol{x}^{*}\boldsymbol{x}^{*}}-K_{\boldsymbol{x}^{*}\boldsymbol{x}}K_{\boldsymbol{x}\boldsymbol{x}}^{-1}K_{\boldsymbol{x}\boldsymbol{x}^{*}}).
\end{align}
The mean of this posterior distribution gives an estimator for our predictions $\vec{f}^{*}$ given the inputs $\vec{\boldsymbol{x}}^{*}$ and the dataset $\{\boldsymbol{x}_{i},f(\boldsymbol{x}_{i})\}$, while the covariance matrix gives an estimation of the accuracy. For our case of interest we choose an exponential kernel~\cite{Varma:2018aht}
\begin{align}
   [K_{\boldsymbol{x}\boldsymbol{x}}]_{ij} =\sigma_{k}^{2}\exp{\left[-\frac{1}{2}\left(\boldsymbol{x}_{i} - \boldsymbol{x}_{j}\right)^{T}\sigma^{-1}\left(\boldsymbol{x}_{i} - \boldsymbol{x}_{j}\right)\right]} + \sigma_{n}^{2}\delta_{ij},
\end{align}
with 
\begin{align}
    \sigma = \text{diag}(\sigma_{1},\dots,\sigma_{D})
\end{align}
and $\sigma_{n}$ a white noise parameter introduced to account for the fact that our training dataset is subject to noise.
The only tunable parameters for GPR are then $\sigma_{k},\sigma,\sigma_{n}$, called hyperparameters. These parameters are determined by maximizing the
marginal likelihood of the training data under the prior multivariate normal distribution~\cite{Varma:2018aht,Wang_2023}. 

This algorithm has been implemented in the \texttt{sklearn} Python
package~\cite{scikit-learn}. Refer to the supplement of Ref.~\cite{Varma:2018aht} for the particular choices used for the Kernel and hyperparameter ranges.

In this work, we trained separate GPR fits for each of the relevant parameters of the BMS transformation. Using the notation defined above, in our case we have
\begin{align}
    \boldsymbol{x}_{i} &= [\ln(q_{i}),(\hat{\chi})_{i},(\chi_{a})_{i}] \\
    f_{i} &= \begin{cases}
                (v_{x})_{i} \\
                (v_{y})_{i} \\
                (v_{z})_{i} \\
                \Re{(\alpha_{\ell m})}_{i} \quad \forall \ell \quad 1 \le \ell \le 8, \ m \ge 0 \\
                \Im{(\alpha_{\ell m})}_{i} \quad \forall \ell \quad 1 \le \ell \le 8, \ m > 0,
            \end{cases}
\end{align}
where the index $i$ labels each simulation in the training dataset characterized in Sec.~\ref{sec:Numerical_evaluation}. For the supertranslation modes we just need to model the $m\ge 0$ modes since $\alpha$ is real. The $\alpha_{00}$ mode is identically set to $0$ in this model. Note that we have not trained the surrogate using the input vector $\boldsymbol{x} = (q,\chi_{1}^{z},\chi_{2}^{z})$. Following Ref.~\cite{Varma:2018aht}, we opted to reparametrize the input space using $(\ln(q), \hat{\chi}, \chi_{a})$ where
\begin{align}
    \hat{\chi} = \frac{\chi_{\mathrm{eff}} - 38\eta\left(\chi_{1}^{z} + \chi_{2}^{z}\right)/113}{1 - 76\eta/113}
\end{align}
with
\begin{align}
    \eta = \frac{q}{\left(1 + q\right)^{2}},
\end{align}
is the spin parameter entering the strain phase at leading-order in the PN expansion~\cite{Khan_2016,Ajith_2011,Cutler_1994,Poisson_1995}.

Furthermore, the dataset has been extended with 13 additional points obtained by symmetry considerations. Indeed, in the case of equal masses $q=1$ and unequal spins, two systems with swapped spins differ only by an overall $\pi$ rotation that has to be applied to both the kick direction and supertranslation parameter modes
\begin{align}
    (v_{x,y})_{i}^{\text{swap}} &= -(v_{x,y})_{i} \\
    (\alpha_{\ell m})_{i}^{\text{swap}} &= \left( -1\right)^{m}(\alpha_{\ell m})_{i} .
\end{align}
\section{Surrogate model error estimation}
\label{sec:errors}
A way of testing the accuracy of the surrogate model is performing a “$k$-fold” cross-validation. We randomly partition our dataset of size $N$ into subsets of $N-k$ elements, then we train a surrogate model on each of these subsets and test it on the points out of the sample. If we denote by $I_{a}^{\mathrm{in}} = \{i_{a_{1}}, \dots i_{a_{N-k}}\}$ the indices of one subgroup $a$, by $I$ the set of all the indices of all the simulations and by $I_{a}^{\mathrm{out}} = I \setminus I_{a}^{\mathrm{in}}$ then
\begin{align}
I &= I_{a}^{\mathrm{out}} \cup I^{\mathrm{in}}_{a} \quad \forall a \\
&= \bigsqcup\limits_{a} I^{\mathrm{out}}_{a} ,
\end{align}
where $\bigsqcup$ denotes disjoint union. Thus, the cross-validation error for each simulation is calculated as
some
\begin{align}
    \Delta\left(f^{*}_{a}(\boldsymbol{x}_{j}),f_{j}\right) \quad \forall j \in I_{a}^{\mathrm{out}}, \ \forall a,
\end{align}
where $f^{*}_{a}(\boldsymbol{x}_{j})$ denotes the prediction of the
surrogate trained on the indices $I_{a}^{\mathrm{in}}$ and tested on a
point $\boldsymbol{x}_{j}$ with index in $I_{a}^{\mathrm{out}}$,
$f_{j}$ is the corresponding value in the training dataset, and
$\Delta$ is some error between the predicted and known values.

In particular, considering the space translation and proper supertranslation parameters we can define
\begin{align}
    \label{eq:transerr}
    \Vert\Delta\alpha_{\ell=1}\Vert &\equiv \Vert(\alpha_{\ell = 1})^{*}_{a}(\boldsymbol{x}_{j}) - (\alpha_{\ell = 1})_{j}\Vert\nonumber\\
    &= \sqrt{\sum_{\vert m\vert \le 1}\vert (\alpha^{*}_{a})_{1 m}(\boldsymbol{x}_{j}) - (\alpha_{j})_{1 m} \vert^{2}} \\
    \label{eq:propsuperr}
    \Vert\Delta\alpha_{\ell\ge 2}\Vert &\equiv \Vert(\alpha_{\ell\ge 2})^{*}_{a}(\boldsymbol{x}_{j}) - (\alpha_{\ell\ge 2})_{j}\Vert\nonumber\\
    &= \sqrt{\sum_{l\ge 2}^{\ell_{\text{max}}}\sum_{\vert m\vert \le \ell}\vert (\alpha^{*}_{a})_{\ell m}(\boldsymbol{x}_{j}) - (\alpha_{j})_{\ell m} \vert^{2}} .
\end{align}
Considering the boost velocities, we estimate a magnitude and an angle error in the following way
\begin{align}
\label{eq:normvelerr}
|\Delta \vec{v}| &\equiv \vert\vec{v}^{*}_{a}\left({\boldsymbol{x}_{j}}\right) - \vec{v}_{j} \vert \\
\label{eq:magvelerr}
|\Delta |\vec{v}|| &= \big\vert\vert\vec{v}^{*}_{a}\left({\boldsymbol{x}_{j}}\right)\vert - \vert\vec{v}_{j} \vert\big\vert \\
\label{eq:angvelerr}
\Delta\theta_{\vec{v}} &\equiv \cos^{-1}{\left(\frac{\vec{v}^{*}_{a}\left({\boldsymbol{x}_{j}}\right) \cdot \vec{v}_{j}}{\vert\vec{v}^{*}_{a}\left({\boldsymbol{x}_{j}}\right)\vert \vert\vec{v}_{j}\vert}\right)}.
\end{align}
We carried out the error estimations by setting $k=5$. 

In the left panel of Fig.~\ref{fig:errorsup} we plot the histogram of the cross-validation relative errors estimated from Eq.~\eqref{eq:propsuperr} over the $L^{2}$ norm of the proper supertranslation parameters in the training dataset and an estimate of the intrinsic numerical relativity resolution error (truncation error) of the simulations. The relative truncation error is obtained by applying Eq.~\eqref{eq:propsuperr} to the training dataset computed from the simulations at the highest resolution and second highest resolution. In the right panel of Fig.~\ref{fig:errorsup} we show the absolute cross-validation errors and truncation errors estimated from Eq.~\eqref{eq:transerr} for the space translation parameters, together with the space translation magnitudes of the training dataset.

Similarly, on the left panel of Fig.~\ref{fig:magangvelerr}, we show the histogram of the cross-validation errors for the norm of the difference [Eq.~\eqref{eq:normvelerr}] and the difference of magnitudes of the boost velocity [Eq.~\eqref{eq:magvelerr}], together with the truncation error. Finally, on the right panel of Fig.~\ref{fig:magangvelerr} we show the histogram relative to the cross-validation angle mismatches [Eq.~\eqref{eq:angvelerr}] together with the corresponding truncation error. 

We note that the cross-validation errors for
$\Vert\Delta\alpha_{\ell=1}\Vert$, $\Vert\Delta\alpha_{\ell\ge 2}\Vert$, $|\Delta
\vec{v}|$, and $\Delta\theta_{\vec{v}}$ are close to the respective truncation errors, indicating that the surrogate correctly captures the features of the dataset and the associated inaccuracies are likely due to numerical resolution errors. We note, though, that the relative errors for $\alpha_{\ell=1}$ are quite high and we suggest using the model with caution for those modes. This could be in part because of the large truncation error, but could also be because of the CoM charge fitting procedure for mapping to the PN frame. The inspiral CoM charge in the PN BMS frame is not perfectly constant, but exhibits small physical oscillations, and thus the use of a linear fit could lead to inaccuracies.

Another possible source of errors could come from not enough accurate PN waveforms and CCE junk radiation in the inspiral portion of the waveform. Nevertheless, with the chosen window parameters ($\gtrsim 500 M$ from the start of the simulation) the initial junk radiation transient is suppressed. Moreover, we did not find significant accuracy improvement by using longer waveforms and mapping to the PN BMS frame at earlier times.
\figerrorsup
\figerrorvel
\section{Comparison of the remnant boost with other kick velocity estimations}
\label{sec:Boostcomparison}
In this project we computed the kick velocity in terms of the BMS boost between the PN BMS frame and remnant black hole BMS frame. In this section, we examine two other ways of computing the boost. 

First, we consider the momentum flux $\frac{d\vec{P}}{dt}$ of gravitational waves defined in Eq.~(2.11) of Ref.~\cite{Ruiz:2007yx}. From the conservation of total momentum~\cite{Gerosa_2018} the remnant kick is\footnote{The exact integration domain should really extend from $u=-\infty$ to $u=+\infty$. Since most of the kick velocity is acquired during the merger, truncating the domain as in Eq.~\eqref{eq:momfluxkick} is a reasonable approximation.}
\begin{align}
\label{eq:momfluxkick}
    \vec{v}_{\text{MF}}(h) \approx -\frac{1}{M_{f}}\int_{u_{\text{PN}}^{*}}^{u_{\text{end}}} \frac{d\vec{P}(h)}{dt} dt ,
\end{align}
where $M_{f}$ is the Christodoulou mass of the remnant, $u^{*}_{\mathrm{PN}}$ is the same time used to frame fix to the PN BMS frame and $u_{\mathrm{end}}$ is the end time of the waveform. Note that to compute the kick from this definition we only need the strain $h$.

Second, we compute the kick by performing a linear fit to the center-of-mass (CoM) charge defined in Eq.~\eqref{eq:comcharge} in the inspiral and post-ringdown, and then taking the difference of each velocity\footnote{Eq.~\eqref{eq:CoMKick} should be formulated in terms of the special relativistic velocity composition rule. Since the kick magnitudes we consider are small compared to $c$, the Newtonian limit is a valid approximation.}
\begin{align}
\label{eq:CoMKick}
v_{\mathrm{CoM}}^{i}  \approx \langle G^{i}(u)\rangle_{\text{Rem}} - \langle G^{i}(u)\rangle_{\text{PN}}.
\end{align}
where $\langle\cdot\rangle$ denotes a linear fit over the inspiral or ringdown windows used for the frame fixing.
This last method is conceptually equivalent to what the BMS transformation method does, as it measures the difference of this BMS charge in the inspiral and remnant portion of the waveform data. 
\figkickcomp
In the right panel of Fig.~\ref{fig:kickcomparison} we can see the comparison between the kick norms obtained from CCE waveforms in the phase fixed PN BMS frame with these alternative calculations.

In Ref.~\cite{Varma:2018aht} the momentum flux kick was computed
through Eq.~\eqref{eq:momfluxkick} on extrapolated
waveforms~\cite{Boyle_2009,Boyle:2019kee}, so we plot such a kick for
comparison. Since the extrapolation procedure provides only the strain, only
the momentum flux method can be used, since the other two require knowledge of the Weyl scalars at $\mathcal{I}^{+}$. Moreover, the extrapolated waveforms have been phase fixed to have the same alignment choice as the CCE ones at $u = -100M$.

As we can see from the left panel of Fig.~\ref{fig:kickcomparison}, the three methods agree when evaluated on the same CCE waveforms, but show a larger absolute discrepancy with the kicks computed from the extrapolated waveforms. Absolute kick discrepancies of this order of magnitude are not new in the literature (see Ref.~\cite{Iozzo:2021vnq}), particularly between extrapolated momentum flux kicks and the kicks obtained from the apparent horizon motion in the simulation domain. However, the relative difference between the kick magnitudes computed from CCE and the extrapolated waveforms remains roughly constant throughout the parameter space, as shown in the right panel of Fig.~\ref{fig:kickcomparison2}, with a median value around $4.4\%$. Such a trend is not observed among other kick discrepancies, indicating a systematic difference between CCE and extrapolated data beyond numerical noise.
\figkickcomptwo

From the right panel of Fig.~\ref{fig:kickcomparison}, we can see that the angle discrepancy between CCE and extrapolated kicks from momentum flux calculations are of the same order as the truncation errors, thus indicating that even though the magnitudes are different, the directions seem to agree.

As Eqs.~\eqref{eq:momfluxkick} and ~\eqref{eq:CoMKick} are BMS covariant, we might expect that a relative boost between CCE and extrapolated waveform could explain this discrepancy. Since the kicks are sufficiently aligned we can try to estimate the boost magnitude in the direction of the kick needed to make the CCE and extrapolated predictions agree. On average, a boost magnitude of $\sim 10^{-3}c$ must be applied to an extrapolated waveform to match the corresponding CCE kick value in either the NR BMS frame or the PN BMS frame. This average value is two orders of magnitude higher than the average boost magnitude that maps the NR BMS frame to the PN frame. Consequently, this hypothesis seems unlikely to explain the discrepancy. However, we remark that a precise BMS frame fix is not possible for the extrapolated waveforms in this dataset, since the extrapolation procedure only provided the strain $h$ and $\Psi_{4}$, thus making it impossible to evaluate all the BMS charges and in particular the CoM charge [Eq.\eqref{eq:comcharge}].
We also note that a partial frame fix has already been done on extrapolated waveforms to account for the gauge-dependent Newtonian center-of-mass drift in the Cauchy domain~\cite{Boyle2016}. However, this does not ensure that the CoM-corrected extrapolated waveform is in any canonical BMS frame, such as the PN one, because the Newtonian center-of-mass is not the same as the PN one and strongly depends on gauge effects in the Cauchy domain.

Additionally, the extrapolated waveforms analyzed are memory corrected through the procedure outlined in Ref.~\cite{Mitman:2020bjf}. This suggests that this discrepancy is likely not due to memory effects, but is sourced by a difference in the multipole moments of the waveforms. To further prove this statement, the green histogram on the left panel of Fig.~\ref{fig:kickcomparison} shows the difference between the kick magnitudes computed from the CCE and extrapolated strains after their energy flux contribution (responsible for displacement memory~\cite{Mitman:2020bjf}) and the angular momentum flux contribution (responsible for the spin memory~\cite{Mitman:2020bjf}) have been removed. As we can see, the discrepancy is unaffected by the removing the flux contributions. We might therefore argue that the extrapolated versus CCE kick discrepancy is sourced by the fundamental difference of these two extraction methods.

\section{Conclusion}
\label{sec:conclusions}
In this work we present the first model for the BMS transformation that maps a binary black hole merger's inspiral frame to the remnant black hole's rest frame. Such a transformation is obtained by composing the BMS transformations that map the BMS charges of the system to their values in the PN BMS frame and remnant black hole BMS frame through the procedure developed in~\cite{Mitman:2021xkq,Mitman:2022kwt}.
The remaining phase degrees of freedom have been fixed by aligning the black holes on the $x$-axis, with the heavier one on the positive axis at a chosen time, using information from the positions of the centers of the apparent horizons from the simulations. 

This procedure has been applied to a dataset containing 89 quasi-circular non-precessing simulations with input parameters $q \le 8$ and $\vert\chi_{1,2}\vert \le 0.8$. From these simulations we built high accuracy fits for the supertranslation and boost parameters using the GPR non-parametric algorithm. 

The fits exhibit a cross-validation error accuracy of $\sim 7 \cdot 10^{-3}M$, (relative, median) for the proper supertranslation sector, of $\sim 3 \cdot 10^{-3}M$ for the space translation sector (absolute, median), and of $\sim 9 \cdot 10^{-3} [0.001c]$ (absolute, median) for the absolute difference of the boost velocity magnitude, and $\sim 0.03$ (median) radians for the angle mismatch, sharing similar values to Ref.~\cite{Varma:2018aht}. The inaccuracies are close to the numerical truncation errors associated to the simulations, therefore suggesting that the fitting technique captures the features of the dataset well, and most of the error comes from the resolution error in the numerical relativity simulations. Nevertheless, we advise for caution when dealing with the space translation parameters fits, as they manifest larger relative errors.

We further explored the comparison between the recoil velocities of the remnant black hole obtained with this model and the ones obtained by other models that use other calculation methods. We found that different methods for computing the kick agree well for CCE waveforms, but those that are obtained through a momentum flux calculation with extrapolated waveforms, as was done in Ref.~\cite{Varma:2018aht}, show a roughly constant relative difference of $\sim 4.4\%$ with respect to those obtained from CCE waveforms. This behavior seems to not be related to frame issues, numerical resolution, or memory effects, but rather due to a genuine differences between the extrapolated and CCE waveforms. More future work will be needed to determine the cause of this systematic discrepancy.

As mentioned previously, this new model extends current fits of remnant parameters by incorporating information about the nonlinear displacement memory, which may be detectable in future gravitational observatories, such as LISA. As a consequence, this model could play a relevant role for GR tests that use the gravitational wave memory. 

Beyond that, this model also has direct applications to waveform modeling. The challenge of creating waveform models from waveforms exhibiting memory lies in the different BMS frames for the inspiral and ringdown. The model proposed in Ref.~\cite{Yoo_2023} handles hybrid PN-CCE waveforms by mapping them to the inspiral BMS frame. However, it struggles to accurately capture ringdown features, likely because the ringdown's BMS frame is distinct across parameter space.

This work proposes a solution by constructing two distinct waveform surrogates: one trained for the inspiral in the inspiral BMS frame and another for the ringdown in the remnant BMS frame. Using our model, these surrogates can then be integrated by mapping the inspiral surrogate predictions to the remnant BMS frame and stitching the two together.

Finally, the proposed model could also aid in estimating the cosmological memory background produced by binary black hole merger populations. 
Current works~\cite{2022SCPMA..6519511Z,Boybeyi_2024,Jokela_2022,zwick2024gravitationalwavememoryimprints} rely on leading-order analytic calculations of memory from either Effective-One-Body (EOB) waveforms or a minimal waveform model~\cite{Favata:2010zu}. Our model instead fully captures nonlinear memory by solely relying on non-perturbative numerical relativity results, thus providing more precise estimates of memory that could be used to produce more accurate models of the cosmological memory background.

The fits produced in this work are available through the Python package \texttt{surfinBH} under the name \texttt{NRSur3dq8BMSRemnant}.
\section*{Acknowledgments}
\label{sec:acknowledgements}
This material is based upon work supported by the National Science Foundation under Grants No.~PHY-2309211, No.~PHY-2309231, No.~OAC-2209656 at Caltech, and No.~PHY-2407742, No.~PHY-2207342, and No.~OAC-2209655 at Cornell. Any opinions, findings and conclusions or recommendations expressed in this material are those of the author(s) and do not necessarily reflect the views of the National Science Foundation. This work was supported by the Sherman Fairchild Foundation at Caltech and Cornell. K.M. is supported by NASA through the NASA Hubble Fellowship grant \#HST-HF2-51562.001-A awarded by the Space Telescope Science Institute, which is operated by the Association of Universities for Research in Astronomy, Incorporated, under NASA contract NAS5-26555.
L.C.S.\ acknowledges support from NSF CAREER Award
PHY–2047382 and a Sloan Foundation Research Fellowship.
S.E.F.~acknowledges support from NSF Grants PHY-2110496 and AST-2407454.
S.E.F.~and V.V.~were supported by UMass Dartmouth's Marine and Undersea
Technology (MUST) research program funded by the Office of Naval Research 
(ONR) under grant no. N00014-23-1-2141.


\bibliography{refs}

\end{document}